\documentclass[epj]{svjour}
\usepackage{graphics}
\begin{document}
\title{Static-light mesons on a dynamical anisotropic lattice}
\author{Justin Foley\inst{1}, Alan \'{o} Cais\inst{2}, Mike Peardon\inst{2}, Sin\'{e}ad M. Ryan\inst{2} 
}                     
%
%
\institute{Department of Physics, Swansea University, Singleton Park, Swansea SA2 8PP, UK  \and School of Mathematics, Trinity College, Dublin 2, Ireland}
\date{Received: date / Revised version: date}
%
\abstract{
We present results for the spectrum of static-light mesons from $\rm{N}_{\rm{f}} = 2$ 
lattice QCD. These results were obtained using all-to-all light quark propagators 
on an anisotropic lattice, yielding an improved signal resolution when compared 
to more conventional lattice techniques. In particular, we consider the inversion 
of orbitally-excited multiplets with respect to the `standard ordering', which has 
been predicted by some quark models.
} 
\maketitle
\section{Introduction}
\label{intro}
It has long been known that hadrons containing a single 
charm or bottom quark exhibit approximate heavy quark spin 
and flavour symmetries and can be described to a reasonable approximation 
by an idealised system in which the heavy quark is taken 
to be infinitely massive~\cite{isgur}. In this limit the heavy quark
symmetries become exact and the system consists of  
light degrees of freedom bound to a static colour point source. 
The static approximation can be systematically improved using a
power-counting rule in $\Lambda_{\rm{QCD}}/{\rm{m}}_{\rm{Q}}$,  which 
forms the basis for heavy quark effective theory (HQET).
For hadrons containing 
a bottom quark this approach is particularly suitable. In this 
case one expects leading-order corrections to be already quite small, at the 
level of ten percent. 
 
On the lattice, mass-dependent errors which arise from the discretisation 
of the Dirac operator can limit the accuracy of simulations of heavy-light
systems. In principle, one can avoid such errors by using the HQET lagrangian to 
describe the heavy quark dynamics. An alternative approach is to perform a number 
of simulations both at unphysically light heavy quark masses using the lattice Dirac operator  
and in the static limit and 
to interpolate the results to the physical heavy quark mass guided by the predictions of HQET.
Therefore, precise calculations in the static limit are of considerable interest to the 
lattice community. 
In addition, little is known experimentally of the excited state spectrum of heavy-light hadrons 
and accurate simulations of their simpler static-light counterparts will have considerable 
phenomenological impact.
However, traditionally, simulations of static-light systems have been extremely noisy. 
Ultimately, this stems from the fact that the use of conventional point-to-all propagators for the light 
 quarks restricts interpolating operators for the static-light hadron to a single 
spatial lattice site. 

This problem could be overcome if it were possible to compute 
all elements of the light quark propagator. 
This would allow 
source and sink operators for the static-light correlators to be placed at every spatial lattice site, yielding 
a dramatic increase in statistics. Indeed, simulations of static-light mesons are the natural 
testing ground for all-to-all techniques since they require a single light quark propagator per configuration while 
the static quark propagator is easy to evaluate. It is therefore not surprising that a number of groups 
have attempted to tackle this particular problem~\cite{green,burch}.

In this contribution, we describe a preliminary study of the excited state spectrum of static-light mesons   
using the all-to-all propagator technique introduced in Ref.~\cite{mise}. 
In this approach, the important contribution of the low-lying eigenmodes of the 
Dirac operator to the fermion propagator is computed exactly. A noisy estimate 
is then used to correct for the contribution of the remaining eigenmodes. 
This estimate is obtained by inverting the Dirac operator on a set of $\rm{Z}_{4}$ noise 
sources which have been partitioned in some subset of quarkfield indices. This partitioning  
or `dilution' yields a substantial reduction in the variance of the stochastic estimate.  
This study has been performed on a 3+1 anisotropic lattice, where the temporal lattice spacing 
$a_{t}$ is much finer than the spatial lattice spacing $a_{s}$. 
The fine temporal lattice spacing proves particularly useful when determining a plateau in the effective 
masses of the higher excited states. 
The correlators have been computed on $\rm{N}_{\rm{f}}=2$ 
background configurations.

\section{Interpolating operators}
\label{sec:1}
Due to heavy quark spin symmetry, states which lie in the same 
hyperfine multiplet are degenerate and in the continuum it is 
conventional to label these multiplets by the angular momentum 
and parity of the light degrees of freedom~\cite{isgur2}. For example, in the static 
limit the pseudoscalar and vector correspond to $ {J_{\ell}}^{P_{\ell}} = {\frac{1} {2}}^{-}$. 

At finite lattice spacing, it makes sense to classify states according to the 
irreducible representations (irreps) of the octahedral point group $\rm{O}_{h}$~\cite{basak1}. This 48 element 
group is the direct product of the 24 proper spatial rotations allowed by the lattice with the group consisting 
of the identity and the space inversion operator. The relationship between these representations 
and continuum quantum numbers is not difficult to determine. Restricting the continuum irreps 
to the elements of the lattice symmetry group generates representations for $\rm{O}_{\rm{h}}$ 
which are in general reducible and identification of the constituent irreps is a straight-forward exercise in group theory. 
Here, we take advantage of the degeneracies which arise from the heavy quark spin symmetry 
and construct interpolating operators specifically for the light degrees of freedom.
Therefore, our operators transform according to the double-valued or spinorial representations of 
${\rm{O}}_{\rm{h}}$.
There are 6 such representations
\begin{eqnarray}
{\rm{G}}_{1 \rm{u}}, {\rm{G}}_{1 \rm{g}}, {\rm{G}}_{2 \rm{u}}, {\rm{G}}_{2 \rm{g}}, {\rm{H}}_{\rm{u}}, 
{\rm{H}}_{\rm{g}}.
\end{eqnarray}
where the subscripts $u$ and $g$ denote representations which 
are odd and even under spatial inversion respectively. 
\begin{table}
\begin{center}
\begin{tabular}{|c|c|c|}
\hline
Lattice irrep & Dimension & ${J}$ \\
\hline
${\rm{G}}_{1}$ & 2 & $\frac{1} {2}, \frac{7} {2}$... \\
\hline
${\rm{G}}_{2}$ & 2 & $\frac{5} {2}, \frac{7} {2}$... \\
\hline
H & 4 & $ \frac{3} {2}, \frac{5} {2}, \frac{7} {2} $..  \\ 
\hline
\end{tabular}
\end{center}
\caption{Irreducible representations of the group of 24 proper spatial 
lattice rotations with low-lying angular momentum content.}
\label{tab:J}
\end{table}

Table~\ref{tab:J} lists the irreps of the subgroup of proper rotations and 
the angular momenta of their lower-lying constituent states.
From this table we see that states with the same angular momentum but with different ${{J}}_{z}$ 
may be scattered across the lattice irreps. 
For example, states which transform according to the rows of the 6-dimensional 
spin $\frac{5} {2}$ representation are divided between the ${\rm{G}}_{2}$ and H irreps.
In a numerical study we should therefore expect to observe energy levels in these irreps which are 
degenerate, up to lattice artifacts, corresponding to the ${J}_{\ell} = \frac{5} {2}$ states.
In this study, we were particularly interested in orbital excitations of the static-light meson. 
There is a single S-wave state  with ${J}_{\ell}^{{P}_{\ell}}$ quantum numbers ${\frac{1} {2}}^{-}$, 
the P-wave doublet is ${\frac{1} {2}  }^{+}, {\frac{3} {2}}^{+}$ and the D-wave states are labelled 
${\frac{3} {2}}^{-}, {\frac{5} {2}}^{-}$. Of these, only the S-wave and the ${\frac{1} {2}}^{+}$ P-wave 
which lie in the $\rm{G}_{1 u}$ and $ \rm{G}_{1 g}$ irreps can be accessed using local operators, while all 
other irreps require the use of extended interpolating operators. It is worth noting that when one uses 
all-to-all light quark propagators the construction 
of extended interpolating fields requires no additional fermion matrix inversions and the extra 
computational cost incurred is minimal.
In order to determine accurately the excited state spectrum it is important to choose operators 
which couple strongly to the states of interest and we tested a number of operators across the 
lattice representations. 
Further details of our choice of interpolating operators will be presented in Ref.~\cite{todhcai}.

\section{Simulation details} 
The lattice Dirac and gauge actions used in this study 
have previously been described in detail in Refs.~\cite{quarkaction,gaugeaction}.  
Both are specifically formulated for 3+1 anisotropic lattices. 
The Wilson-like fermion action is accurate to $\mathcal{O}(a_{t}, a_{s}^{3})$ and employs 
stout-smeared spatial links to reduce radiative corrections.
The gauge action incorporates tree-level Symanzik and tadpole improvement and is accurate to
$\mathcal{O} ( a_{t}^{2}, a_{s}^{4}, \alpha_{s} a_{s}^{2} )$.

The anisotropic lattice breaks hypercubic symmetry and introduces two new parameters: 
the bare quark and gauge anisotropies, denoted $\xi_{q}^{0}$ and $\xi_{g}^{0}$, which must be tuned such that the measured anisotropy, $a_{s}/a_{t}$,  assumes a consistent value. In dynamical QCD the tuning procedure is quite complicated and is 
detailed in Ref.~\cite{morrin}. In this study we use bare anisotropy values taken from that paper 
of $\xi_{q}^{0} = 7.52$ and $\xi_{g}^{0} = 8.06$ which yields a measured anisotropy of 6 with a 
three percent error.

We employ the standard Eichten-Hill action~\cite{eichten} for the static quark and the corresponding 
static quark propagator is a product of unsmeared temporal links.

Measurements have been performed on 249 background configurations on an $8^{3} \times 80$ lattice 
with a spatial lattice spacing of about $0.2\rm{fm}$.  
The mass of the sea quarks and the light valence quark was found to be 
approximately the strange quark mass. For the light quark propagator, we 
computed the contribution of 100 low-lying eigenvectors of the 
Dirac operator. Two independent $\rm{Z}_{4}$ noise sources which were 
diluted in time and colour were used in the stochastic estimate. 

To extract excited state energies we applied the standard variational 
analysis to correlation matrices which were obtained by applying 5 or 6 
levels of Jacobi smearing to the light quark fields. This was facilitated 
by the use of all-to-all propagators which meant that no additional 
inversions were required to smear the quark fields.

\section{Results}

\begin{figure}
\resizebox{0.48\textwidth}{!}{%
  \includegraphics*{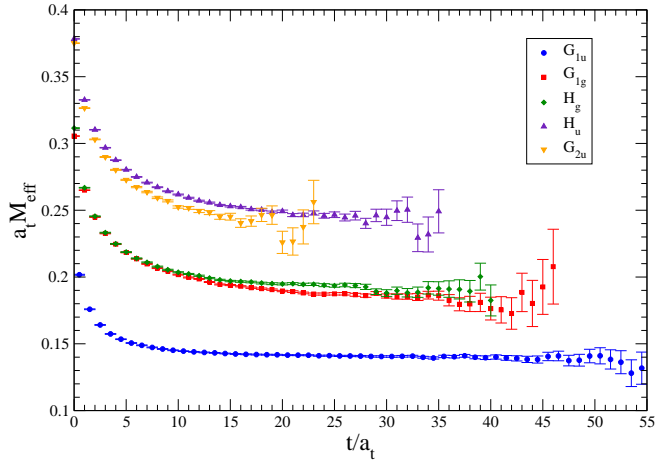}
}
\caption{Effective masses for the lowest-lying states in 5 lattice irreps. These include P-wave and D-wave excitations.}
\label{all}       
\end{figure}

Fig.~\ref{all} shows effective masses for the lowest-lying states across a number of the 
lattice irreps. In this case the lowest effective mass corresponds to the S-wave ${\frac{1} {2}}^{-}$ state
and the higher-lying states are orbital excitations. The highest-lying states lie in D-wave channels. 
Clear plateaux are evident in each channel. 

\begin{figure}
\resizebox{0.48\textwidth}{!}{
  \includegraphics*{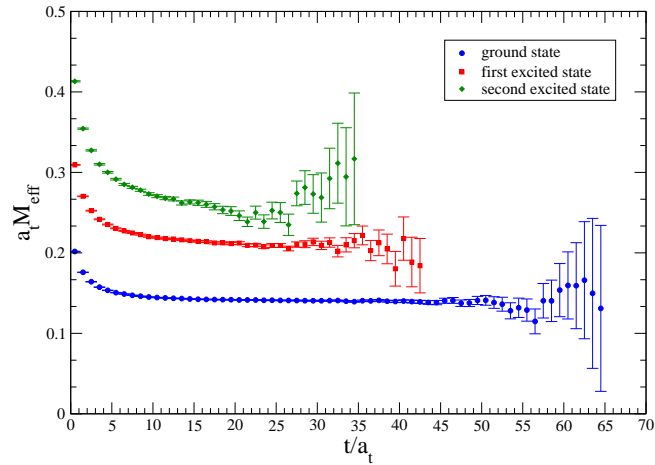}
}
\caption{Effective masses for the lowest-lying states in the $G_{1 u}$ (S-wave) irrep.}
\label{swave}
\end{figure}

In addition, we obtain strong signals for a number states 
in each of the lattice irreps. Fig.~\ref{swave} shows the ground state and the 
first two excited states  
in the $\rm{G}_{1 u}$ (S-wave) irrep. 
These excited states shown here
are almost certainly radial excitations in the S-wave channel.

\begin{figure}
\resizebox{0.48\textwidth}{!}{
  \includegraphics*{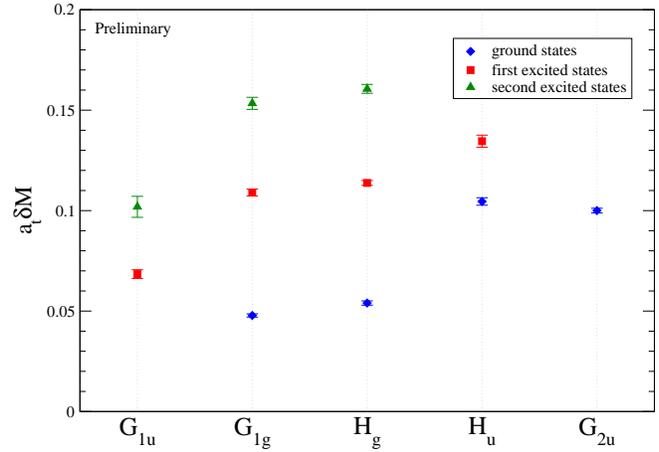}
}
\caption{Mass differences between the ground state S-wave multiplet
and higher lying states.}
\label{splitting}
\end{figure}

Since we are working in the static limit the energies determined 
by fitting to the effective masses contain an unphysical shift. 
This shift cancels in energy differences, which are therefore physically 
meaningful. Fig.~\ref{splitting} plots the energy differences between the lowest-lying 
state in the $\rm{G}_{1 u}$ representation (i.e. the S-wave multiplet)  and a number of 
excited states. We have not presented these preliminary results in physical units 
because we have performed our simulation on quite a small spatial volume which 
might distort the energies of the orbitally and radially excited 
states.  
The main point here is the precision to which we can compute the spectrum. 
Nevertheless, it is interesting to note the position of the lowest-lying states in the 
$\rm{H}_{u}$ and $\rm{G}_{2 u}$ irreps. Naively, one expects the lightest state in the $\rm{H}_{u}$ representation 
to correspond to the ${\frac{3} {2}}^{-}$ D-wave while the ${\frac{5} {2}}^{-}$ ought to be the lightest 
state in the $\rm{G}_{2 u}$ irrep. However, there appears to be no significant splitting between 
these states which may indicate that the ${\frac{5} {2}}^{-}$ multiplet 
is lighter than the ${\frac{3} {2}}^{-}$. Such a scenario has been predicted 
by quark models~\cite{schnitzer2,isgur3}.

%

\section{Conclusions and outlook}
We have presented preliminary results for the spectrum 
of static-light mesons in $\rm{N}_{\rm{f}} = 2$ QCD. We have been 
able to compute energy differences between the S-wave 
multiplet and a number of orbital and radial excitations. 
The use of all-to-all propagators allows us to minimise the 
statistical uncertainty in our measurements and facilitates the 
use of spatially-extended interpolating operators and the 
variational approach. 
The results presented here have allowed 
us to make some tentative remarks about the nature of the spectrum;
however, we have not yet made a serious 
attempt to identify the continuum quantum numbers of the observed states. 
More recently we have 
performed similar simulations using larger operator bases on 
two different spatial volumes. These runs 
should allow us to assess finite volume effects and identify 
multi-particle states.  
We are currently analysing the results of these 
simulations and we intend to present our findings in the near future~\cite{todhcai}.

\end{document}